\documentclass[intlimits,twoside,a4paper]{article}
\usepackage{graphicx}
\usepackage{subcaption} 
\usepackage{float}
\usepackage{appendix}
\usepackage{listings}
\usepackage{xcolor}
\usepackage[cp1251]{inputenc}

\usepackage[eqsecnum]{cmpj3}

\DeclareMathOperator{\erf}{erf}

\issue{2025}{28}{3}{33701}
\doinumber{10.5488/CMP.28.33701}

\title[Generalised two-dimensional nonlinear oscillator]%
{Generalised two-dimensional nonlinear oscillator with a position-dependent effective mass and the thermodynamic properties}%

\author[S. E. Bokpe, F. A. Dossa, G. Y. H. Avossevou]
{S. E. Bokpe\orcid{0009-0006-4939-6941}\refaddr{label1}, F. A. Dossa \orcid{0000-0002-2694-4144}\refaddr{label2}\thanks{Corresponding author: \email{dossafanselme@gmail.com}.},
 G. Y. H. Avossevou\orcid{0000-0002-9609-0340}\refaddr{label1}}

\addresses{
\addr{label1} Institut de Math\'ematiques et de Sciences  Physiques (IMSP),
 Universit\'e d'Abomey-Calavi (UAC),
01 BP 613 Porto-Novo, R\'epublique du B\'enin
\addr{label2} Laboratory of Physics and Applications (LPA), Universit\'e Nationale des Sciences, Technologies,
Ing\'enierie et Math\'ematiques (UNSTIM) Abomey, BP: 2282 Goho Abomey, R\'epublique du B\'enin
}


\date{Received January 4, 2025, in final form May 30, 2025}

\begin{document}
\maketitle
\begin{abstract}

We investigate a two-dimensional nonlinear oscillator with a position-dependent effective mass in the framework of nonrelativistic quantum mechanics. Using the Nikiforov-Uvarov method, we obtain exact analytical expressions for the energy spectrum and wave functions. Based on the canonical partition function, we derive key thermodynamic quantities, including internal energy, specific heat, free energy, and entropy. Our results show that, unlike the one-dimensional case, where the specific heat is unaffected by the nonlinearity parameter $k$, the two-dimensional system exhibits a strong $k-$dependence. At high temperatures, the specific heat becomes temperature-independent for fixed values of $k$, in line with the \emph{Dulong--Petit} law. However, these behaviors occur only for negative values of $k$. These findings highlight the impact of effective mass nonlinearity on macroscopic thermodynamic quantities and suggest that tuning the parameter $k$ could serve as an effective strategy for enhancing the performance of quantum devices, including thermal machines and optoelectronic components.

\printkeywords
%
\end{abstract}

\section{Introduction}

In quantum mechanics, nonlinear systems are obtained by studying quantum problems modeled by a quantum oscillator on three-dimensional spherical and hyperbolic spaces\cite{a1,a2,a3,a4} or in curved space in general. Others are considered as variable mass systems with enormous interests due to their various applications in different fields of physics such as semiconductor fabrication \cite{1}, quantum liquids \cite{2}, helium clusters \cite{3}, the study of electronic properties of inhomogeneous crystals. In fact, some particles such as free electrons can have a different mass than the electrons that evolve in crystals. This mass varies according to the position occupied by the electron in the crystal and its sign depends on that of the nonlinearity parameter. This nonlinearity parameter can be positive or negative \cite{4,5,6}. From a geometric point of view, the study of variable mass systems in two dimensions proves that they are not only integral but super-integral. This implies mathematical difficulties because of the kinetic energy term comprising a position-dependent mass \cite{1,7, 8, 20, 21}. In \cite{8}, it is shown that a nonlinear system has periodic solutions whose frequency depends on the amplitude.
 It has also been proven that there is a conjugation relation between this nonlinear system and the linear harmonic oscillator on spaces of constant curvature, the two-dimensional sphere S$^{2}$ and the hyperbolic plane H$^{2}$. In \cite{9}, it is shown that variable mass systems treated in one dimension exhibit the form invariance \cite{24} and realize finite-dimensional Lie algebras such as the Heisenberg--Weyl algebra, $su(1,1)$ and $su(2)$. Other studies have been carried out on effective mass systems such as the supersymmetric approach to coherent states in the sense of Barut and Girardelo~\cite{10}, the approach using transport and dispersion properties of heterojunctions \cite{11} and the Schr\"odinger factorization method approach \cite{7,9,10,22,23,arc}.

Indeed, it is important to determine the thermodynamic properties of a harmonic oscillator with effective mass because of its uses in statistical thermodynamics, quantum mechanics, and solid-state physics. These properties allow us to model physical systems such as phonons in a crystal lattice or electrons in a semiconductor. In such a lattice, the vibration of atoms around their equilibrium position effects the heat capacity of the material. The knowledge of the thermodynamic properties of a harmonic oscillator with position-dependent mass helps optimize optoelectronic devices where the thermal response of electrons impacts the behavior of lasers, diodes, and transistors. Motivated by these applications, we decided to study the thermodynamic properties of a harmonic oscillator with variable mass using the Nikiforov-Uvarov method. This method is very useful in solving the Schr\"odinger equation for variable-mass systems.
Recently, in \cite{31}, the method is used to study the Schr\"odinger equation with a general ambiguity, a position dependent mass Morse
potential. More recently, in \cite{32}, this method is used to
search for the exact one-dimensional solutions of the position
 dependent mass Schr\"odinger equation with pseudoharmonic oscillator and its thermal properties thermodynamic properties.
 However, the two-dimensional of the thermodynamic properties of variable mass systems has not been done.
 
  In this work, we proposed to study the two-dimensional quantum model of a variable mass system using the well-known techniques of the Nikiforov-Uvarov method \cite{12,13,14,15}. We analytically determine the wave function and the energy spectrum of the system. Then, we evaluate the partition function from which we study the behavior of the thermodynamic properties of the system.
Our study reveals that the effect of the nonlinearity parameter on the thermodynamic properties is considerable. It has been shown that in one dimension \cite{10} the specific heat does not depend on the nonlinearity parameter but in our work in two dimensions we find that the specific heat is effected by the nonlinearity parameter $k$ and remains insensitive at high temperature.
This result is also considerable since the specific heat is a function of the structure of a substance.

The work is organized as follows: in section~\ref{sec2}, we define a nonlinear system whose mass depends on the position. Then, section~\ref{sec3} is devoted to the determination of the wave function and the energy spectrum from the Hamiltonian of the two-dimensional system. In section~\ref{sec4}, we calculate the partition function and then  different thermodynamic functions such as average energy,  heat capacity,  free energy and entropy. Finally, the last section contains a conclusions.

\section{Two-dimensional nonlinear oscillator}\label{sec2}

The study of the nonlinear oscillator problem is of great importance in classical and quantum mechanics because the problem plays a leading role in the explanation of a large number of realistic physical phenomena. Thus, a nonlinear bi-dimensional oscillator is a two-dimensional system that is described by a nonlinear differential equation. It can be studied using the Lagrangian and Hamiltonian formalisms.
It was recently shown in \cite{8} that there is a generalization to $n$ dimensions preserving the symmetry characteristics.
This is the only generalization to $n$ dimensions for which the kinetic term is a quadratic function in the velocities, which is invariant under rotations and under the
two vector fields generalizing the symmetries of the one dimensional model \cite{7, 10}
In particular, the two-dimensional generalization studied in \cite{8} which was not only integrable but also superintegrable is given by the following Lagrangian,
\begin{eqnarray}\label{La}
	L=\frac{m(r)}{2}\left[v_r^2+k(xv_y-yv_x)^2\right]-\frac{1}{2}m(r)\alpha^2 r^2,\quad v_r^2=v_x^2+v_y^2,\quad  r^2=x^2+y^2,
\end{eqnarray}
where $k={\delta^{2}}/{\lambda}$ represents the nonlinearity parameter on which the mass depends.
$\alpha$ is the frequency of the oscillations while
$m(r)$ represents the variable mass of the system.
%
There are several variable mass models in the literature\cite{9}. In this work, we choose the variable mass whose expression is given by
\begin{eqnarray}
m(r)=\frac{\lambda}{1+\delta^{2} r^{2}},
\end{eqnarray}
with $\lambda$ being a real parameter and $\delta$ a constant that measures the force of oscillator linearity.

The Hamiltonian corresponding to the Lagrangian (\ref{La}) is given by
\begin{eqnarray}
	H=\frac{(1+\delta^{2}r^{2})}{2\lambda} (P_{x}^{2}+P_{y}^{2})-\frac{\delta^{2}}{2\lambda}(xP_{y}-yP_{x})^{2} +\frac{\alpha^{2}\lambda r^{2}}{2\left(1+\delta^{2} r^{2}\right)}. \label{h}
	\end{eqnarray}
	It should be noted that the $k-$dependence is introduced in two distinct ways: the first is through the global factor $ {(1+\delta^{2}r^{2})}/{\lambda}$, which is the most direct extension to $n=2$ of the
one-dimensional factor ${(1+\delta^{2}x^{2})}/{\lambda}$; the second is term $\frac{\delta^{2}}{2\lambda}(xP_{y}-yP_{x})^{2}$, which represents a genuinely two-dimensional contribution that is absent in the one-dimmensional case. This additional term considerably effects the properties of the system.
	
Now, we can use the polar coordinate system. Thus, we have
\begin{equation}
x=r\cos\theta,\quad y=r\sin\theta,\quad r^2=x^2+y^2,
\end{equation}
\begin{equation}
p_x^2+p_y^2= \dfrac{\partial^{2}}{\partial r^{2}}+\frac{1}{r}\dfrac{\partial}{\partial r}-\frac{1}{r^{2}}L^2_z,\quad L_{z}^{2}=-\dfrac{\partial^{2}}{\partial \theta^{2}}.
\end{equation}
The eigenvalue equation of the Hamiltonian $(\ref{h})$ is given by
 \begin{eqnarray}
 \left[-\frac{(1+\delta^{2} r^{2})}{2\lambda}\frac{\partial^{2}}{\partial r^{2}}-\frac{(1+\delta^{2} r^{2})}{2r\lambda}\frac{\partial}{\partial r}+ \frac{L_z^2}{2\lambda r^{2}} +\frac{\alpha^{2}\lambda r^{2}}{2\left(1+\delta^{2} r^{2}\right)}\right]\varPsi\left(r,\theta \right)=E\varPsi\left(r\theta \right),\label{EE}
\end{eqnarray}
with $\hbar=1$. We can write the wave function $\varPsi(r, \theta)$ as follows
\begin{eqnarray}
 \varPsi\left(r,\theta \right)=U(r)\re^{-\ri m\theta},\label{E}
 \end{eqnarray}
 with $m=0,\pm1,\pm2,\ldots$. The above equation becomes,
 \begin{eqnarray}
 \left[-\frac{(1+\delta^{2} r^{2})}{2\lambda}\frac{\partial^{2}}{\partial r^{2}}-\frac{(1+\delta^{2} r^{2})}{2r\lambda}\frac{\partial}{\partial r}- \frac{m^2}{2\lambda r^{2}} +\frac{\alpha^{2}\lambda r^{2}}{2\left(1+\delta^{2} r^{2}\right)}\right]U(r)=E U(r).\label{EE}
\end{eqnarray}
To solve this equation, we use the Nikiforov-Uvarov method which is one of the analytical methods used to solve the Schr\"odinger equation that resists standard
methods. Its differential equation generalized hyperbolic type \cite{26}\cite{27} is presented in the form:
 \begin{eqnarray}
 \varPsi^{\prime\prime}(z)+\dfrac{\tilde{\tau}(z)}{\sigma(z)}\varPsi^{\prime}(z)+\frac{\tilde{\sigma}(z)}{\sigma^{2}(z)}\varPsi(z)=0\label{1}.
 \end{eqnarray}
 The most convenient parametric form of $\eqref{1}$ \cite{13} is written
 \begin{eqnarray}
 \frac{\rd^{2}\varPsi(z)}{\rd z^{2}}+\dfrac{a_{1}-a_{2}z}{z(1-a_{3}z)}\frac{\rd\varPsi(z)  }{\rd z}+\dfrac{1}{z^{2}(1-a_{3}z)^{2}}\left[ -\epsilon_{1} z^{2} + \epsilon_{2} z- \epsilon_{3}\right]\varPsi(z)=0, \label{3}
 \end{eqnarray}
 with
 \begin{eqnarray}
 \tilde{\tau}(z)=a_{1}-a_{2}z,\quad\sigma(z)=z(1-a_{3}z),\quad\tilde{\sigma}(z)=-\epsilon_{1} z^{2} + \epsilon_{2} z- \epsilon_{3},
 \end{eqnarray}
 $\tilde{\tau}$ is a polynomial of degree not greater than 1. $\sigma(z)$ and $\tilde{\sigma}(z)$ are polynomials of degree not greater than~$2$.
 
 The functions $ \pi(z) $ and the parameter $\kappa\pm$ are defined by:
 \begin{eqnarray}
 \pi(z)=a_{4}+a_{5}z\pm\sqrt{(a_{6}-\kappa a_{3})z^{2}+(a_{7}+\kappa)z+a_{8}},
 \label{p}
 \end{eqnarray}
 \begin{eqnarray}
\kappa\pm=-(a_{7}+2a_{3}a_{8})\pm2\sqrt{a_{8}a_{9}}
\label{pi}.
 \end{eqnarray}
 For the method to be valid, the function
 \begin{eqnarray}
 \tau(z)=\tilde{\tau}(z)+2\pi(z),
 \end{eqnarray} 
  would have to satisfy the condition that its derivate must be negative.  This condition is met when using $\kappa_{-}$ expression. In this approach, we find different solutions of the system.

The weight function  $ \rho(z) $, from Nikiforov-Uvarov method, is given by the following relation:
 \begin{eqnarray}
 \rho(z)=z^{a_{10}-1}(1-a_{3}z)^{\frac{a_{11}}{a_{3}}-a_{10}-1},\label{l} 
 \end{eqnarray}
 and
 \begin{eqnarray}
 y_{n}(z)=P_{n}^{\left( a_{10}-1,\frac{a_{11}}{a_{3}}\right)}(1-2a_{3}z).\label{l1}
 \end{eqnarray}
  Part of the general function
  \begin{eqnarray}
 P_{n}^{(a,b)}(1-2z)=\dfrac{\Gamma(n+a+1)}{n!\Gamma(a+1)}{}_{2}F_{1}(-n,1+n+a+b,1+a,z),
  \end{eqnarray}
  represents the Jacobi polynomial \cite{28}.
The second part of the general function is given by:
 \begin{eqnarray}
 \phi(z)=z^{a_{12}}(1-a_{3}z)^{-\frac{a_{13}}{a_{3}}-a_{12}},\label{l2}
 \end{eqnarray}
The general solution $ \chi(z)=\phi(z) y_{n}(z)$ is then
\begin{eqnarray}
\chi(z)=z^{a_{12}}(1-a_{3}z)^{-\frac{a_{13}}{a_{3}}-a_{12}}P_{n}^{\left( a_{10}-1,\frac{a_{11}}{a_{3}}\right)}(1-2a_{3}z).
\end{eqnarray}
  We use the following relation to explicitly determine the energy spectrum of the system:
 \begin{eqnarray}
 a_{2}n-(2n+1)a_{5}+n(n-1)a_{3}+(2n+1)(a_{3}\sqrt{a_{8}}+\sqrt{a_{9}})+a_{7}+2a_{3}a_{8}+2\sqrt{a_{8}a_{9}}=0,\label{ne}
 \end{eqnarray}
 with the different parameters~\cite{13,25}
 \begin{eqnarray}
 a_{4}&=&\frac{1}{2}(1-a_{1}),\quad a_{5}=\frac{1}{2}(a_{2}-2a_{3}),\quad a_{6}=a_{5}^{2}+\epsilon_{1},\nonumber\\
 a_{7}&=&2a_{4}a_{5}-\epsilon_{2},\quad a_{8}=a_{4}^{2}+\epsilon_{3},\quad a_{9}=a_{3}a_{7}+a_{3}^{2}a_{8}+a_{6},\nonumber\\
 a_{10}&=&a_{1}+2a_{4}+2\sqrt{a_{8}},\quad a_{11}=a_{2}-2a_{5}+2(\sqrt{a_{9}}+a_{3}\sqrt{a_{8}}),\nonumber\\
 a_{12}&=&a_{4}+\sqrt{a_{8}},\quad a_{13}=a_{5}-(\sqrt{a_{9}}+a_{3}\sqrt{a_{8}}).
 \end{eqnarray}
 The radial differential equation is given in the form
 \begin{eqnarray}
 \frac{\rd^{2}U(r)}{\rd r^{2}}+\frac{1}{r}\frac{\rd U(r)}{\rd r}+\left[\frac{2\lambda E}{1+\delta^{2} r^{2}}-\frac{m^{2}}{r^{2}(1+\delta^{2} r^{2})}-\frac{\alpha^{2}\lambda^{2} r^{2}}{(1+\delta^{2} r^{2})^{2}} \right]U(r) =0.  \label{n}
 \end{eqnarray}

 \section{Spectre of the system}\label{sec3}

 The NU method is based on solving a second-order linear differential equation by reducing it to a generalized equation of the hypergeometric type. Using the appropriate coordinate transformation $ z=-\delta^{2} r^{2} $, this equation can be rewritten in the following form:
 \begin{eqnarray}
 \frac{\rd^{2}U(z)}{\rd z^{2}}+\dfrac{1-z}{z(1-z)}\frac{\rd U(z)}{\rd z}+\dfrac{1}{z^{2}(1-z)^{2}}\left( \mu z^{2} + \gamma z- \omega\right)U(z)=0,
 \end{eqnarray}
 with
\begin{eqnarray}
\mu =\frac{\lambda E}{2\delta^{2}}-\frac{\alpha^{2}\lambda^{2}}{4\delta^{4}},\quad\gamma = \frac{m^{2}}{4}-\frac{\lambda E}{2\delta^{2}},\quad
\omega=\frac{m^{2}}{4}.
\end{eqnarray}
 By identification, we find
 \begin{eqnarray}
 a_{1}=a_{2}=a_{3}=1,\quad\epsilon_{1}=-\mu,\quad \epsilon_{2}=\gamma,\quad\epsilon_{3}=\omega.
 \end{eqnarray}
  The expressions for the other coefficients are as follows:
\begin{eqnarray}
a_{4}&=&0,\quad a_{5}=-\frac{1}{2},\quad a_{6}=\frac{1}{4}-\mu,\quad a_{7}=-\gamma,\quad a_{8}=\omega,\nonumber\\
a_{9}&=& \omega - \gamma-\mu+\frac{1}{4},\quad a_{10}=1+2\sqrt{\omega },\nonumber\\
a_{11}&=&2+2\left(\sqrt{ \omega -\gamma-\mu+\frac{1}{4}}+\sqrt{\omega}\right),\quad
a_{12}=\sqrt{\omega },\nonumber\\
a_{13}&=& -\frac{1}{2}-\left(\sqrt{ \omega - \gamma-\mu+\frac{1}{4}}+\sqrt{\omega}\right).
\end{eqnarray}
 From the relations $\eqref{p}$ and $ \eqref{pi}$ we find
 \begin{eqnarray}
 	\pi(z)=- \frac{1}{2}z\pm\sqrt{\Big(\frac{1}{4}-\mu-\kappa \Big)z^{2}+(\kappa-\gamma)z+\frac{m^{2}}{4} },\\
 	\kappa\pm=-\Big(\frac{\lambda E}{2\delta^{2}}+\frac{m^{2}}{4}\Big)\pm\frac{\lvert  m\rvert}{2}\sqrt{\frac{\alpha^{2}\lambda^{2}}{\delta^{4}}+1}.
 \end{eqnarray}
 The physically acceptable expression for $\pi(z)$ is the one for which
 \begin{eqnarray}
 	\tau(z)=\tilde{\tau}(z)+2\pi(z),
 \end{eqnarray} has a negative derivative. Using the expression for $\kappa_{-}$, the appropriate forms of $\pi(z)$ and $ \tau(z)$ are given respectively by
 	\begin{eqnarray}
 	\pi(z)=-\frac{1}{2}\left[\left( 1+\lvert  m\rvert+\sqrt{\frac{\alpha^{2}\lambda^{2}}{\delta^{4}}+1}\right) z-\lvert  m\rvert\right] ,\\
 	\tau(z)=-\left( 2+\sqrt{\frac{\alpha^{2}\lambda^{2}}{\delta^{4}}+1}+\lvert  m\rvert\right) z+1+\lvert  m\rvert.
 	\end{eqnarray}
 	The derivative of $ \tau(z) $ gives us
 	\begin{eqnarray}
 	\tau'=-\left( 2+\lvert  m\rvert+\sqrt{\frac{\alpha^{2}\lambda^{2}}{\delta^{4}}+1}\right)<0.
 	\end{eqnarray}
The relation (\ref{l}) allows us to write the weight function in the form
 	\begin{eqnarray}
 	\rho(r)=(-\delta^{2}r^{2})^{\lvert  m\rvert}(1+\delta^{2} r^{2})^{\sqrt{\frac{\alpha^{2}\lambda^{2}}{\delta^{4}}+1}}.
 	\end{eqnarray}
 	The relations (\ref{l1}), (\ref{l2}) allow us to write respectively
 \begin{eqnarray}
 y_{n}(r)=P_{n}^{\left( \lvert  m\rvert,\sqrt{\frac{\alpha^{2}\lambda^{2}}{\delta^{4}}+1}\right) }(1+2\delta^{2}r^{2}),
 \end{eqnarray}
 \begin{eqnarray}
 \phi(r)=(-\delta^{2} r^{2})^{\frac{\lvert  m\rvert}{2}}(1+\delta^{2} r^{2})^{\frac{1}{2}+\frac{1}{2}\sqrt{\frac{\alpha^{2}\lambda^{2}}{\delta^{4}}+1}},
 \end{eqnarray}
 with
 \begin{align}
 P_{n}^{\left( \lvert  m\rvert,\sqrt{\frac{\alpha^{2}\lambda^{2}}{\delta^{4}}+1}\right) }(1+2\delta^{2}r^{2})&=\dfrac{\Gamma(n+\lvert  m\rvert+1)}{n!\Gamma(\lvert  m\rvert+1)}\nonumber\\
 &\times{}_{2}F_{1}\left( -n,1+n+\lvert  m\rvert+\sqrt{\frac{\alpha^{2}\lambda^{2}}{\delta^{4}}+1},1+\lvert  m\rvert,-\delta^{2} r^{2}\right).
\end{align}
The general radial solution $ U(r)=\phi(r)y_{}(r)$ is then
\begin{eqnarray}
 U(r)=(-\delta^{2} r^{2})^{\frac{\lvert  m\rvert}{2}}(1+\delta^{2} r^{2})^{\frac{1}{2}+\frac{1}{2}\sqrt{\frac{\alpha^{2}\lambda^{2}}{\delta^{4}}+1}}P_{n}^{\left( \lvert  m\rvert,\sqrt{\frac{\alpha^{2}\lambda^{2}}{\delta^{4}}+1}\right) }(1+2\delta^{2}r^{2}).
\end{eqnarray}
The total wave function is then written as
\begin{eqnarray}
\varPsi(r,\theta)=C_{n,m}(-\delta^{2} r^{2})^{\frac{\lvert  m\rvert}{2}}(1+\delta^{2} r^{2})^{\frac{1}{2}+\frac{1}{2}\sqrt{\frac{\alpha^{2}\lambda^{2}}{\delta^{4}}+1}}P_{n}^{\bigg( \lvert  m\rvert,\sqrt{\frac{\alpha^{2}\lambda^{2}}{\delta^{4}}+1}\bigg) }\big(1+2\delta^{2}r^{2}\big)\re^{\ri m\theta },
\end{eqnarray}
with $C_{n,m}$ the normalization constant.

Using the relation $\eqref{ne}$ and setting $k={\delta^{2}}/{\lambda}$ we can write the expression of the energy spectrum of the system in the form
  \begin{eqnarray}
 E_{n_{r},m}= \left(2n_{r}+\lvert  m\rvert+1 \right)\sqrt{\alpha^{2}+k^{2}}-k\left[2n_{r}^{2}+\frac{m^{2}}{2}+\left(2n_{r}+1 \right)\left(\lvert  m\rvert+1 \right)   \right],
 \end{eqnarray}
with $k<0$.

 \section{Thermodynamic properties}\label{sec4}

Let us now study the thermodynamic properties of our system. Thus, the partition function is calculated to determine the thermodynamic functions, namely: internal energy, heat capacity, free energy and entropy.
The partition function can be expressed as follows \cite{13,16}
 \begin{eqnarray}
 Z=\sum_{n=0}^{\lambda}\re^{-\beta E_{n}},
 \end{eqnarray}
 with $\lambda$ being the upper bound, $k_{\text{B}} $ represents the Boltzmann constant and $\beta={1}/{k_{\text{B}}T}$. 
 
Using the Poisson summation given by \cite{13,17}:
 \begin{eqnarray}
 \sum_{n=0}^{N}f(n)=\frac{1}{2}\big[f(0)-f(N+1) \big]+\int_{0}^{N+1}f(x)\rd x.
 \end{eqnarray}
 We find
 \begin{eqnarray}
  Z=\frac{1}{2}\big(\re^{\beta(\tilde{a}-\tilde{d})}-\re^{-\beta\tilde{c}}\big)-\Omega,
  \end{eqnarray}
with
\begin{eqnarray}
\tilde{a}&=&k(\lvert  m\rvert+1)-\sqrt{\alpha^{2}+k^{2}}, \quad
\tilde{b}=-(3+\lvert  m\rvert+2\lambda)k+\sqrt{\alpha^{2}+k^{2}}, \nonumber\\
\tilde{c}&=&(2\lambda+\lvert  m\rvert+3)\sqrt{k^{2}+\alpha^{2}}-k\left(2\lambda^{2}+\frac{m^{2}}{2}+6\lambda+5+2\lambda \lvert  m\rvert+3\lvert m\rvert\right),\nonumber\\ \tilde{d}&=&\lvert  m\rvert\sqrt{\lambda^{2}+\alpha^{2}}-\frac{\lambda  m^{2}}{2},
\quad\Omega=\frac{1}{2k}\sqrt{\frac{\piup}{2}}\re^{-\frac{\alpha^{2}\beta}{2k}}\left( \frac{\tilde{a}\erf( \sqrt{\eta}) }{\sqrt{2\eta}}+\frac{\tilde{b} \erf( \sqrt{\vartheta}) }{\sqrt{2\vartheta}} \right),\nonumber \\  \vartheta&=&-\dfrac{\beta\tilde{b}^{2}}{2k}, \quad \eta=-\dfrac{\beta\tilde{a}^{2}}{2k},\quad
 \erf(z)=\frac{2}{\sqrt{\piup}}\int_{0}^{z}\re^{-t^{2}}\rd t.
 \end{eqnarray}

 {\bf Average energy $U$:}

 \begin{eqnarray}
 U=-\frac{\partial}{\partial \beta}\ln Z
  =-\frac{\Lambda}{\big(\re^{\beta(\tilde{a}-\tilde{d})}-\re^{-\beta\tilde{c}}\big)-2\Omega  },
 \end{eqnarray}
 where
  \begin{eqnarray} \Lambda=(\tilde{a}-\tilde{b})\re^{\beta(\tilde{a}-\tilde{b})}+\tilde{c}\re^{-\beta\tilde{c}}+\frac{(\alpha^{2}\beta+k)\Omega
 }{k\beta}-\frac{1}{2k\beta}\re^{-\frac{\alpha^{2}\beta}{2k}}\Big(\tilde{a}\re^{\frac{\tilde{a}^{2}\beta}{2k}}+\tilde{b}\re^{\frac{\tilde{b}^{2}\beta}{2k}}\Big).
\end{eqnarray}

 {\bf Heat capacity:}

 \begin{eqnarray}
 	C=-k_{\text{B}}\beta^{2}\frac{\partial U}{\partial \beta},
 \end{eqnarray}
  \begin{eqnarray}
 	C=\frac{1}{2}k_{\text{B}}\beta^{2}\left[ \frac{(\tilde{a}-\tilde{b})^{2}\re^{\beta(\tilde{a}-\tilde{b})}-\tilde{c}^{2}\re^{-\beta\tilde{c}}-\varepsilon}{[\re^{\beta(\tilde{a}-\tilde{d})}-\re^{-\beta\tilde{c}}]-2\Omega }-\frac{2\Lambda^{2}}{([\re^{\beta(\tilde{a}-\tilde{d})}-\re^{-\beta\tilde{c}}]-2\Omega)^{2} }\right],
  \end{eqnarray}
 where \begin{eqnarray}
 \varepsilon&=&-\frac{1}{2k^{2}\beta^{2}}(\alpha^{4}\beta^{2}+2\alpha^{2}\beta k+3k^{2})\Omega-\frac{\re^{-\frac{\alpha^{2}\beta}{2k}}}{4k^{2}\beta^{2}}\varsigma,\nonumber
 \\\varsigma&=&\tilde{a}\re^{\frac{\tilde{a}^{2}\beta}{2k}}(\tilde{a}^{2}\beta-2\alpha^{2}\beta-3k)+\tilde{b}\re^{\frac{\tilde{b}^{2}\beta}{2k}}(\tilde{b}^{2}\beta-2\alpha^{2}\beta-3k).
  \end{eqnarray}

 {\bf  Free energy:}

 \begin{eqnarray}
 F= -\frac{\ln Z}{\beta}
 =\frac{\ln2-\ln\left(\re^{\beta(\tilde{a}-\tilde{d})}-\re^{-\beta\tilde{c}}-2\Omega   \right)}{\beta}.
 \end{eqnarray}

\begin{figure}[h]
	\centering
	\includegraphics[width=0.65\linewidth]{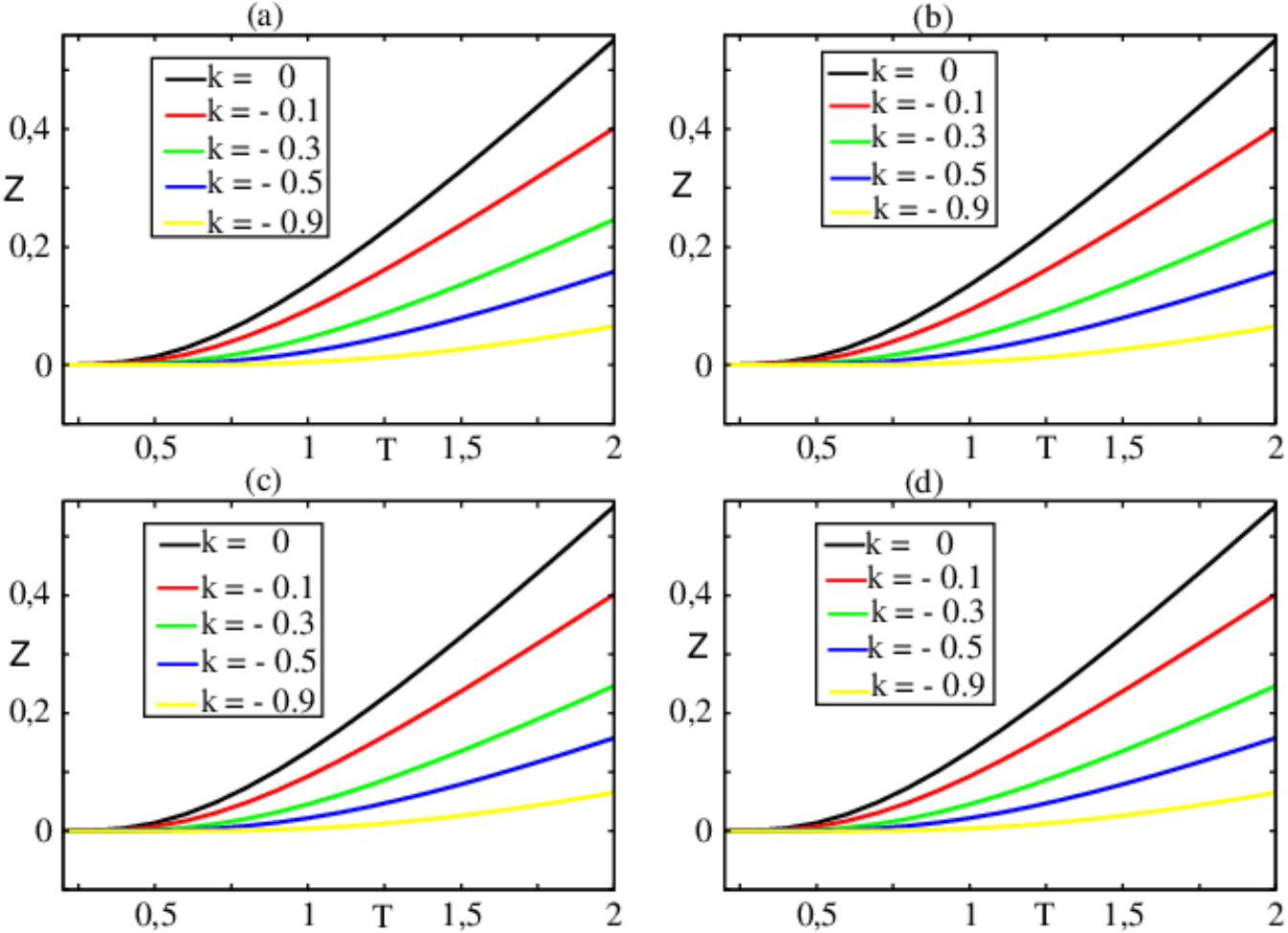}
	\caption{(Colour online) 
		Partition function as a function of temperature for different values of the parameter $k$
		In (a): $ \lambda=200$, in (b) $ \lambda=300$, in (c) $ \lambda=400$, in (d) $ \lambda=500$.
	}
	\label{fig:par}
\end{figure}

 {\bf Entropy:}
\begin{eqnarray}
S= k_{\text{B}}\beta^{2}\frac{\partial F}{\partial \beta},
 \end{eqnarray}
 \begin{eqnarray}
S=k_{\text{B}}\left[ -\ln2+\ln\left( \re^{\beta(\tilde{a}-\tilde{d})}-\re^{-\beta\tilde{c}}-2\Omega   \right) - \frac{\beta\Lambda}{[\re^{\beta(\tilde{a}-\tilde{d})}-\re^{-\beta\tilde{c}}]-2\Omega  }\right].
\end{eqnarray}

\begin{figure}[h]
	\centering
	\includegraphics[width=0.7\linewidth]{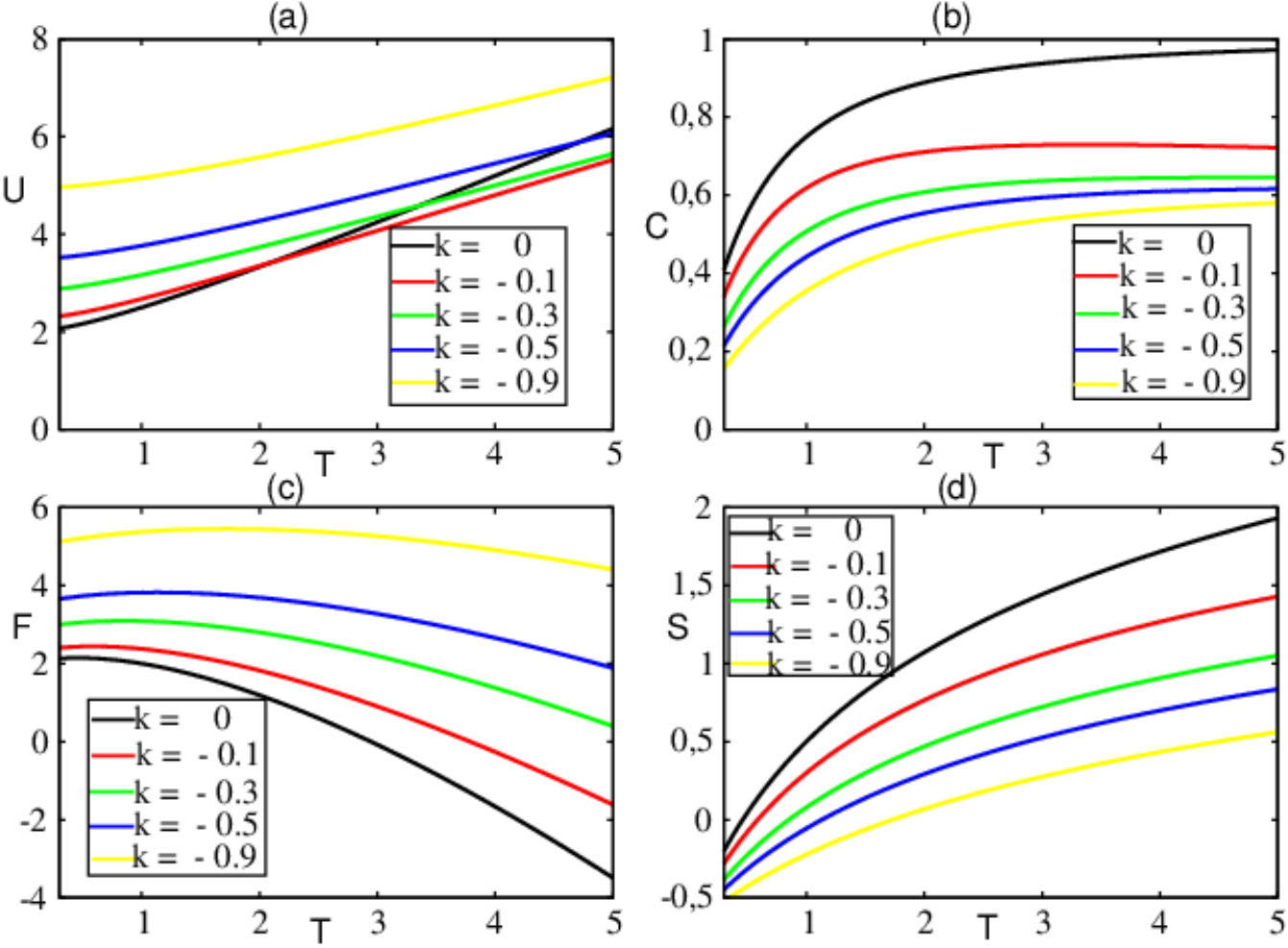}
	\caption{(Colour online) (a) average energy, (b) heat capacity, (c) free energy, (d) entropy as a function of $T$ for different values of parameter $k$; $m=1$ , $k_{\text{B}}=1$ and $\lambda=500$.}
\label{fig:ener1}
\end{figure}

\begin{figure}[h]
	\centering
	\includegraphics[width=0.7\linewidth]{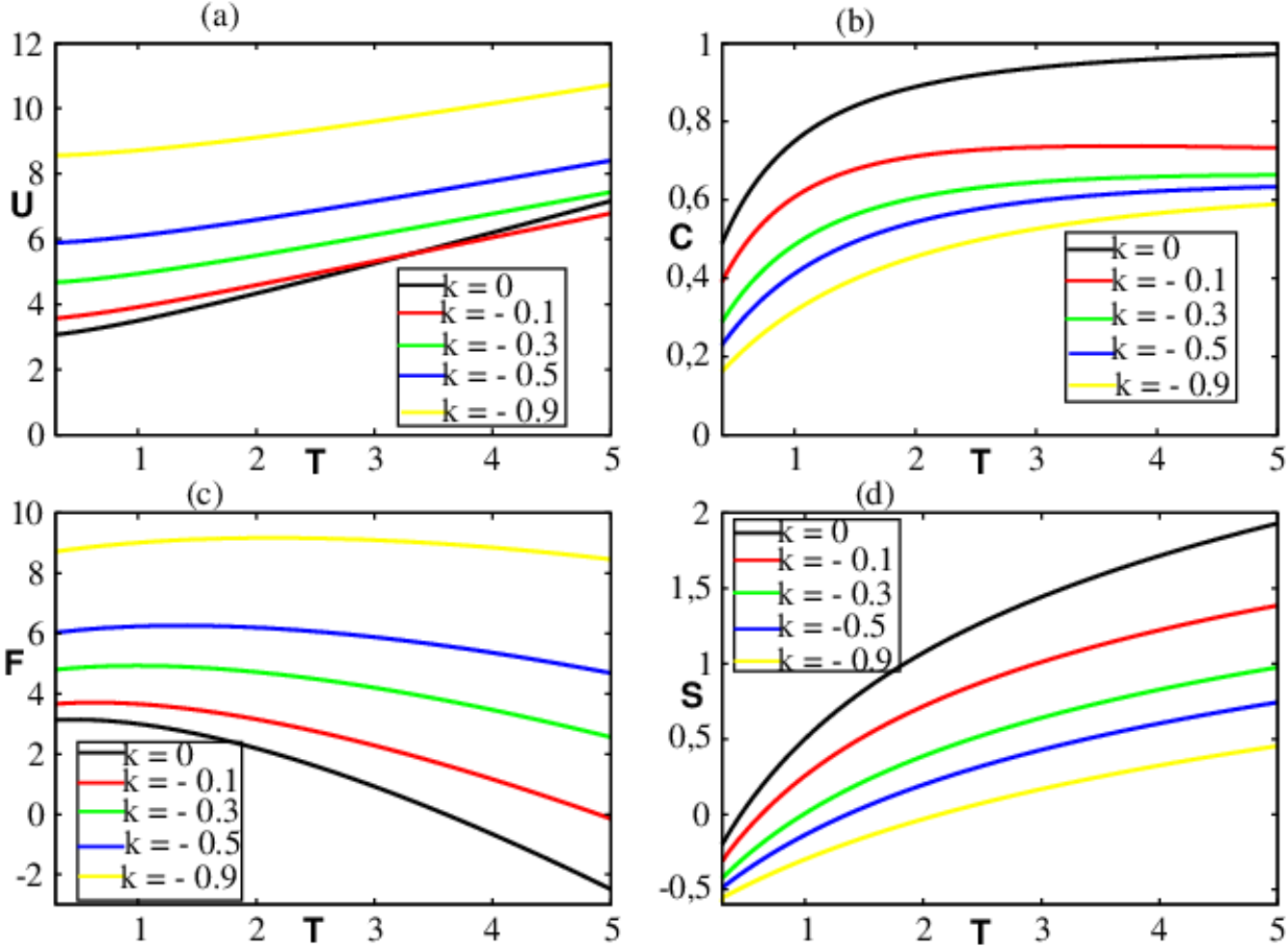}
	\caption{(Colour online) (a) average energy, (b) heat capacity, (c) free energy, (d) entropy as a function of $T$ for different values of parameter $k$; $m=2$ , $k_{\text{B}}=1$ and $\lambda=500$.}
	\label{fig:ener2}
\end{figure}

The partition function and thermodynamic function curves are plotted as
a function of temperature for different values of the non-lineraty parameter
$k$.
In figure $\ref{fig:par}$, we see that the partition functions increase monotonously with temperature for each
fixed value of the nonlinearity parameter $k$. The curves are insensitive from $\lambda=300$. For clarity, we have used the larger $\lambda=500$ to plot all thermodynamic functions.
In figures $\ref{fig:ener1}$ and $\ref{fig:ener2}$, we note that the average energy increases as the temperature increases. The specific heat increases rapidly for low temperatures and then becomes an almost constant value at high temperatures. The free energy decreases while the entropy increases as the temperature increases. We also note that the effect of the parameter $k$ is more apparent at the level of the average energy for lower temperatures but the effect is felt at the level of the specific heat, free energy and entropy for higher temperatures.

We note that, in \cite{arc}, the authors focused on the construction of new exactly solvable one-dimensional potentials by deforming shape-invariant potentials, highlighting theoretical effects such as the noncommutativity between mass and momentum operators, as well as a modified Hermiticity condition. Their analysis reveals considerable structural modifications in the energy spectra and potential shapes as a function of the mass parameters, with fundamental implications in quantum mechanics. In our study, we investigated a two-dimensional nonlinear oscillator with position-dependent effective mass, applying the analytical Nikiforov-Uvarov method in order to rigorously determine the wave functions, energy spectra, and thermodynamic properties of the system. Our results show that the nonlinearity parameter directly effects the thermodynamic functions, particularly the specific heat, which may have practical applications in the optimization of quantum devices such as thermal machines and optoelectronic components.

\section{Conclusions}

In this work, the Nikiforov-Uvarov method has been applied to study a two-dimensional nonlinear oscillator with a position-dependent effective mass. This approach has enabled us to exactly determine  the energy spectrum and the corresponding wave functions. From these results, the canonical partition function has been evaluated, allowing the derivation of key thermodynamic quantities such as the mean energy, specific heat, free energy, and entropy.

Our analysis has shown that, for each fixed value of the nonlinearity parameter $k$, the mean energy and entropy increase with temperature, while the free energy decreases. The specific heat has been observed to increase with temperature before reaching a saturation value, consistent with the {Dulong--Petit } law. Notably, this constant value decreases as the parameter $k$ decreases, highlighting a direct effect of nonlinearity on the thermal behavior of the system. A similar dependence on $k$ has been found in the behavior of entropy.

These findings are in agreement with the previous one-dimensional studies, but reveal a key distinction: in two dimensions, the specific heat depends on the nonlinearity parameter $k$, unlike in the one-dimensional case. This result is particularly significant, as it illustrates how a macroscopic thermodynamic quantity can be strongly affected by the quantum characteristics of the system. Furthermore, we have found that meaningful thermodynamic behavior is obtained only for certain negative values of $k$, suggesting that tuning the effective mass via this parameter may offer a promising route for optimizing quantum devices, such as thermal machines and optoelectronic components.

\ukrainianpart

\title{Термодинамічні властивості узагальненого двовимірного нелінійного осцилятора із залежною від положення ефективною масою}
\author{С. Е. Бокпе\refaddr{label1}, Ф. А. Досса\refaddr{label2},
	Г. Й. Х. Авоссеву\refaddr{label1}}
\addresses{
	\addr{label1} Інститут математики та фізичних наук,
	Університет Абомей-Калаві, 01 BP 613 Порто-Ново, Бенін
	\addr{label2}  Лабораторія прикладної фізики, Національний університет наук, технологій, інженерії та математики (UNSTIM) Абомей, BP: 2282 Гохо Абомей, Бенін}
%
%
%
\newpage
\makeukrtitle

\begin{abstract}
	\tolerance=3000%
	Досліджується двовимірний нелінійний осцилятор із залежною від положення ефективною масою в рамках нерелятивістської квантової механіки. Використовуючи метод Нікіфорова-Уварова, ми отримуємо точні аналітичні вирази для енергетичного спектру та хвильових функцій. На основі канонічної функції розподілу отримано ключові термодинамічні величини включно з внутрішньою енергію, питомою теплоємністю, вільною енергією та ентропією. Отримані результати показують, що, на відміну від одновимірного випадку, коли параметр нелінійності $k$ не впливає на питому теплоємність, двовимірна система демонструє сильну $k$-залежність. За високих температур питома теплоємність стає незалежною від температури для фіксованих значень $k$, що відповідає закону \emph{Дюлонга--Пті}. Однак така поведінка спостерігається лише для від'ємних значень $k$. Ці результати підкреслюють вплив нелінійності ефективної маси на макроскопічні термодинамічні величини та свідчать про те, що налаштування параметра $k$ може служити ефективною стратегією для підвищення продуктивності квантових пристроїв, включаючи теплові машини та оптоелектронні компоненти.
	\keywords нелінійний осцилятор Нікіфорова-Уварова, термодинамічні властивості
	
\end{abstract}

\lastpage
\end{document}